\documentclass[review]{elsarticle}

\usepackage{amsmath}
\usepackage{amsfonts}
\usepackage{amssymb}
\usepackage{adjustbox}
\usepackage{hyperref}
\usepackage{comment}
\usepackage{pdflscape}
\usepackage{caption}
\usepackage{hyperref}
\usepackage{xcolor}
\definecolor{myorange}{rgb}{0.9921, 0.5529, 0.2352}
\definecolor{my300}{RGB}{158, 154, 200}
\definecolor{my600}{RGB}{27, 158, 119}

\usepackage{graphicx}
\usepackage{subcaption}

\usepackage{lineno,hyperref}
\modulolinenumbers[5]

\journal{}

\biboptions{authoryear}

\begin{document}

\begin{frontmatter}

\title{Cryptocurrency Dynamics: Rodeo or Ascot?}

\author[mymainaddress]{Konstantin H\"ausler}
\author[mymainaddress,mysecondaryaddress]{Wolfgang Karl H\"ardle}

\nonumnote{Corresponding Author: Konstantin H\"ausler, haeuslek@hu-berlin.de,  Humboldt Universit\"at zu Berlin, Germany.}

\nonumnote{ Financial support of the  European Union's Horizon 2020 research and innovation program "FIN-TECH: A Financial supervision and Technology compliance training programme" under the grant agreement No 825215 (Topic: ICT-35-2018, Type of action: CSA), the European Cooperation in Science \& Technology COST Action grant CA19130 - Fintech and Artificial Intelligence in Finance - Towards a transparent financial industry, the Deutsche Forschungsgemeinschaft's IRTG 1792 grant, the Yushan Scholar Program of Taiwan, the Czech Science Foundation's grant no. 19-28231X / CAS: XDA 23020303 are greatly acknowledged.}

\address[mymainaddress]{Humboldt-Universit\"at zu Berlin, Germany}

\address[mysecondaryaddress]{Wang Yanan Institute for Studies in Economics, Xiamen University, China. \\ Sim Kee Boon Institute for Financial Economics, Singapore Management University. \\ Faculty of Mathematics and Physics, Charles University, Czech Republic.\\ National Chiao Tung University, Taiwan.}

\begin{abstract}
We model the dynamics of the cryptocurrency (CC) asset class via a stochastic
volatility with correlated jumps (SVCJ) model with rolling-window parameter
estimates. By analyzing the time-series of parameters, stylized patterns are
observable which are robust to changes of the window size and supported by
cluster analysis. During bullish periods, volatility stabilizes at low levels and
the size and volatility of jumps in mean decreases. In bearish periods though,
volatility increases and takes longer to return to its long-run trend. Furthermore, jumps in mean and jumps in volatility are independent. With the rise of the CC market in 2017, a level shift of the volatility of volatility occurred. All codes are available on  \href{https://github.com/QuantLet/SVCJrw}{\includegraphics[height=.9em]{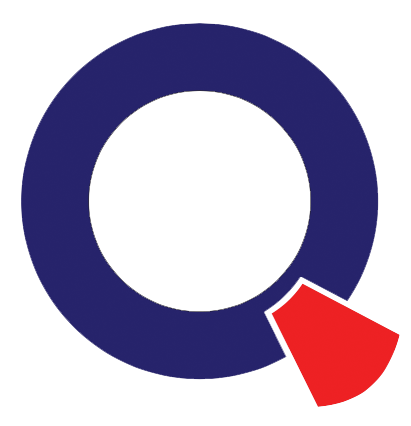} \color{magenta}Quantlet.com}
\end{abstract}


\end{frontmatter}


\section{Introduction}
The rise of the cryptocurrency (CC) asset class opens up just as many opportunities as it raises questions. In particular, the functioning of its dynamics is not yet understood:
first attempts to characterize this sector by standard time series methods had limited success (cf. \cite{shichen2016}). 
This paper investigates the dynamics of the CC market by modeling them via a stochastic volatility with correlated jumps (SVCJ) model in combination with rolling window parameter estimates. Thereby we obtain time series for each parameter, and these time series reveal several recurring patterns, which we interpret as stylized facts.

There exists a large literature that analyzes sub-areas of the CC sector. Several indices track its dynamics (e.g. \cite{trimborn2018}, \cite{elendnerf5}, \cite{cci30})
and several characteristics have been identified: \cite{stylized2018} report heavy tails of the return distributions of CCs, the cointegration relationships of the top CCs by market capitalization is highlighted by  \cite{keilbar2021} and the high volatility compared to classical assets is emphasized by \cite{hardle2020}, just to name a few. However, understanding the big picture is still a challenge. \cite{chen_cathy2018} show that an SVCJ model can meet the challenges of explaining the dynamics of such a non-stationary sector. This paper takes up and extends their method by a rolling window approach to broaden the view on the CC sector and to shed light on its dynamics. The SVCJ model, introduced by \cite{duffie2000}, assumes a stochastic movement of the index returns as well as a stochastic movement of their volatility. Also, co-jumps of prices and volatility are considered. We shift several rolling windows of differing sizes through the data, and at each time step, we estimate the parameters of the SVCJ model in a Bayesian manner. Thereby we obtain time series for each parameter, which allow to characterize the dynamics of the CC sector.
\\
In general, parameter estimates are time-varying and sensitive to the window size. However, several recurring patterns are observable that are robust to changes in window size and are supported by k-means clustering of the parameter estimates: First, volatility remains at a low level during bullish CC market movements and rises in times of bearish markets. Besides, when volatility is already on a high level, it needs longer to return to its long-run trend. Second, in times of bullish markets, the size of jumps in mean return decreases, and its volatility stabilizes as well at low levels. Third, a level shift of the volatility of volatility parameter occurred simultaneously to the rise of the CC market at the turn of the year 2017/18. 
Thus, while the dynamics of the CC sector are not orderly like a horse race at Ascot, they are not a wild rodeo ride either.
\\

The remainder of this paper is structured as follows: Section \ref{data} introduces the CRIX, a CC index that is used as a representative of the CC sector in the following analysis. Section \ref{svcj_model} explains the methodology and estimation approach and Section \ref{svcj_estimates} presents the estimation results and their robustness checks. Section \ref{cluster_analysis} reveals the dependencies among parameter estimates by $k$-means clustering and thereby identifies several stylized facts on the CC market dynamics. All codes are available on  \href{https://github.com/QuantLet/SVCJrw}{\includegraphics[height=.9em]{qletlogo_tr.png} \color{magenta}Quantlet.com}


\section{Data}{\label{data}}

We use the CRIX, a CC index that tracks the price dynamics of the whole CC sector, developed at the Blockchain Research Center at Humboldt University Berlin by \cite{trimborn2018}. By optimally adjusting the number of constituents dynamically, the CRIX ensures high accuracy in reflecting the CC market dynamics. The CRIX data is obtained from \href{thecrix.de}{thecrix.de} and the whole sample period in this paper is from January, 2015 to July, 2020.

\begin{figure}[h!]
\includegraphics[width=\textwidth]{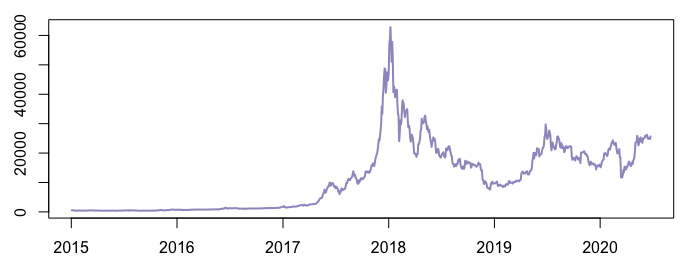}
\caption{The dynamics of CRIX from January, 2015 to July, 2020. Data source: \href{thecrix.de}{thecrix.de}}
\label{CRIX_plot}
\end{figure}

\section{Methodology}{\label{chapter_dynamics}}

\cite{shichen2016} have shown that standard econometric time series methods  like ARIMA-GARCH processes cannot capture the dynamics of the non-stationary CC market. \cite{chen_cathy2018} however catch the dynamics of the CC market very accurately by using an SVCJ model firstly introduced by \cite{duffie2000}. In this setting, the CRIX index value can be modeled by the following Euler discretization version of the SVCJ model:

\begin{subequations}
\begin{align}
Y_{t} &=\mu+\sqrt{V_{t-1}} \varepsilon_{t}^{y}+Z_{t}^{y} J_{t} \label{mu} \\
V_{t} &=\alpha+\beta V_{t-1}+\sigma_{v} \sqrt{V_{t-1}} \varepsilon_{t}^{v}+Z_{t}^{v} J_{t}
\label{empirical_calibration}
\end{align}
\end{subequations}
where $Y_{t+1}=\log \left(S_{t+1} / S_{t}\right)$ denotes the daily log return of the CRIX index and $\mu$ is the trend term. $\varepsilon_{t}^{y}$ and $\varepsilon_{t}^{v}$ are two standard normal r.v.s that are correlated at rate $\rho$. $\alpha$ is the drift of the volatility and $\beta$ is related to the speed of mean-reversion. $\sigma_v$ denotes the volatility of the volatility parameter. The jump process is modeled by jump size $Z_t$ and jumpf frequency $J_t$: 
$Z_{t}^{v}$, the size of jumps in volatility, follows an exponential distribution with parameter $\mu_{v}$ and conditional on it, $Z_{t}^{y}$ (the size of jumps in returns) follows a normal distribution with mean $\mu_{y}+\rho_{j} Z_{t}^{v}$ and variance $\sigma_{y}^{2}$. $J_{t}$ is a Bernoulli r.v. with $P \left(J_{t}=1\right)=\lambda$. \\

We run the implementation procedures developed by \cite{perez2018} to calibrate the parameter family $\{\mu, \mu_y, \sigma_y, \lambda, \alpha, \beta, \rho, \sigma_v, \rho_j, \mu_v\}$. The code is available on  \href{https://github.com/QuantLet/SVCJOptionApp/tree/master/}{\protect\includegraphics[height=.9em]{qletlogo_tr.png} \color{magenta} Quantlet.com}. 
Specifically, each parameter at time $t$ is estimated with data from the window $[t-n, t-1]$, where $n$ is the window size.
We estimate the model parameters in a rolling-window way of the sample periods and analyze the time-series patterns of the parameters in different settings of window size.

\section{Empirical Results}
\subsection{Dynamics of the Cryptocurrency Market}{\label{svcj_estimates}}

\begin{figure}[h!]
\begin{subfigure}{\textwidth}
\includegraphics[width = 1.1\textwidth]{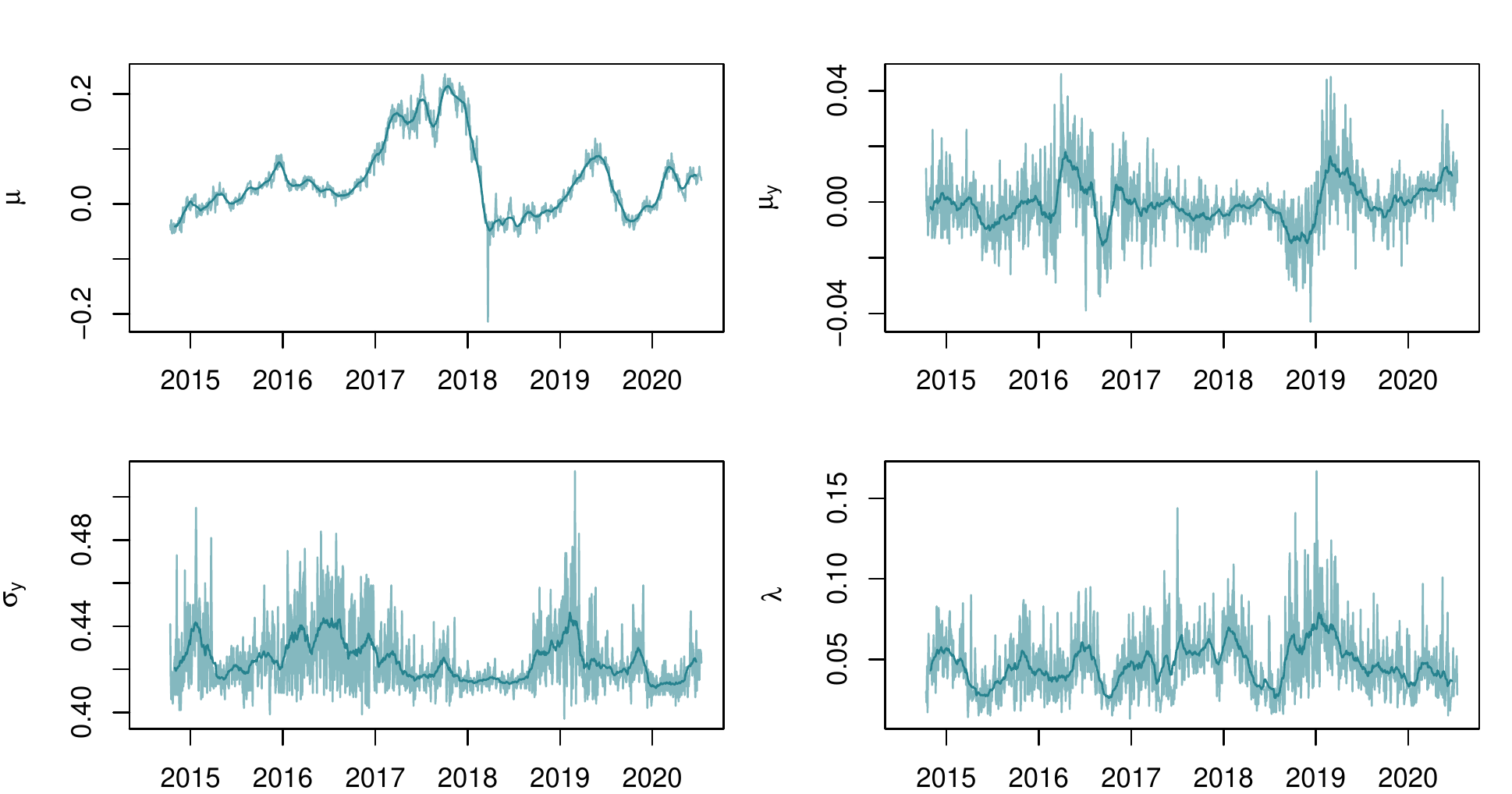}
\includegraphics[width = 1.1\textwidth]{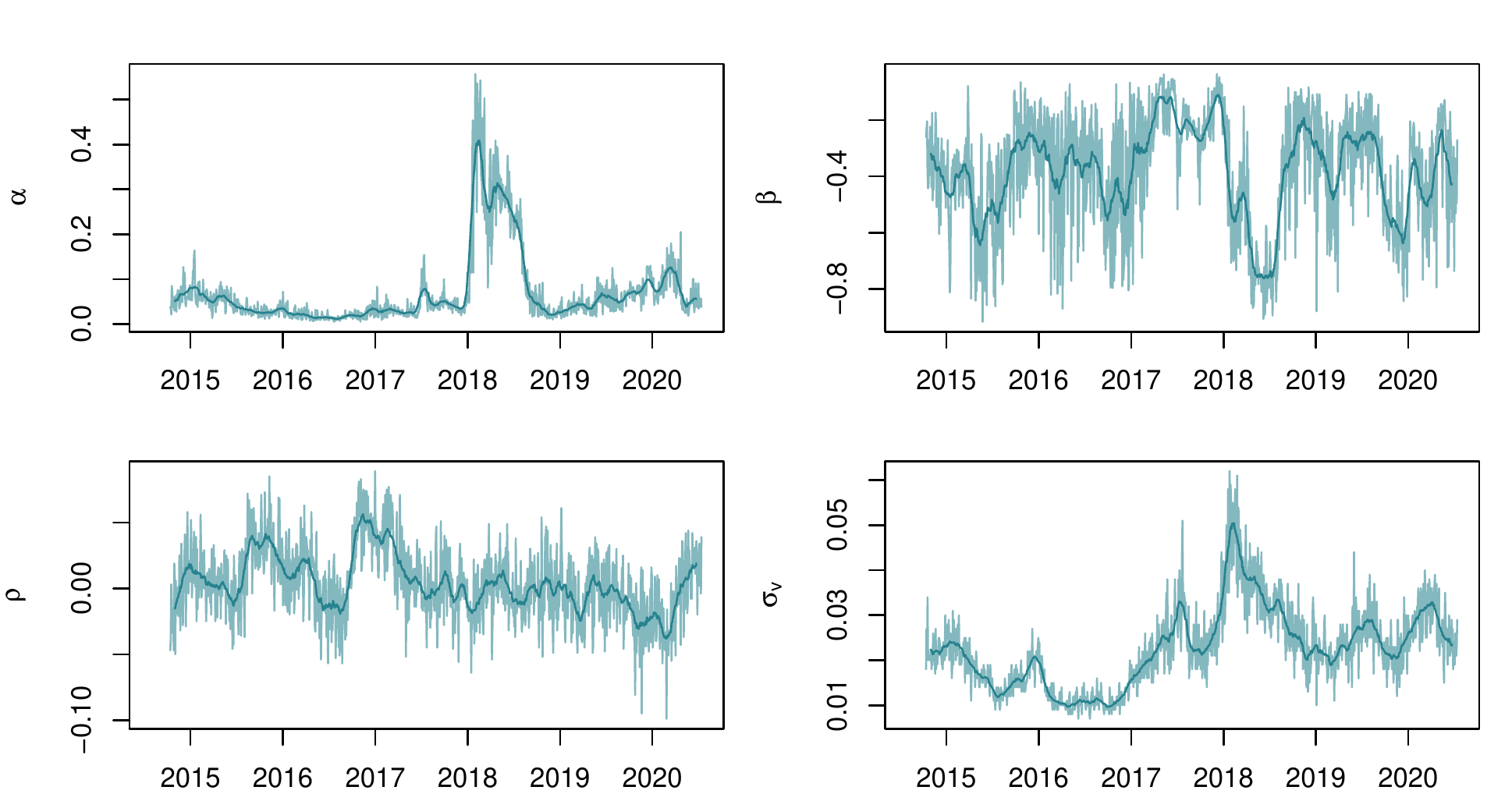}
\includegraphics[width = 1.1\textwidth]{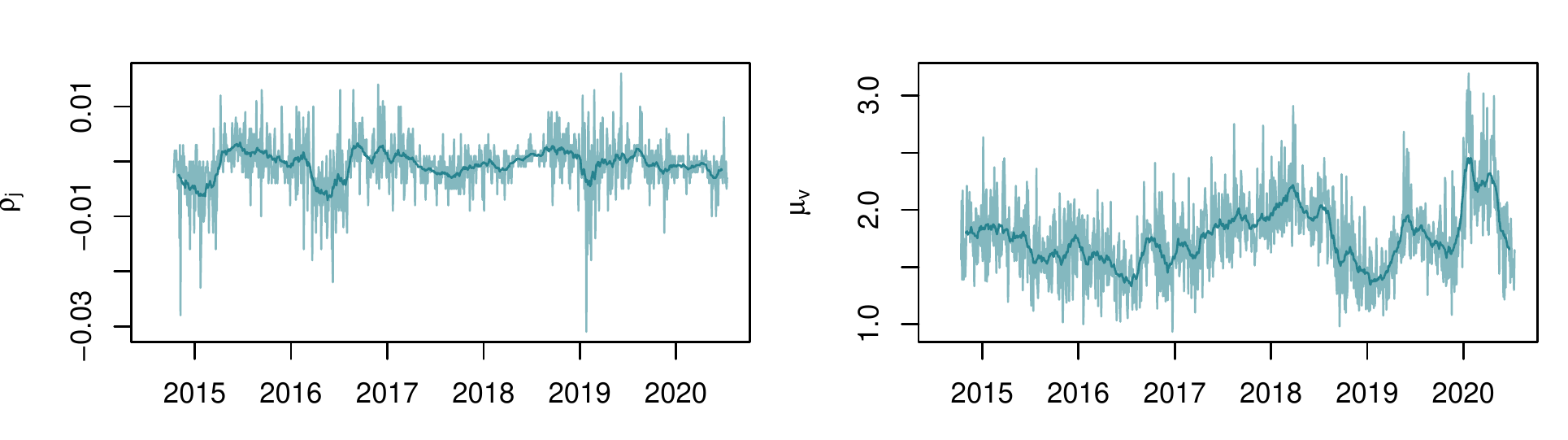}
\label{fig:B}
\end{subfigure}
\caption{Parameter estimates of the SVCJ model with a rolling window of 150 days. The fluctuating lines represent actual parameter estimates, the solid lines in their center depict moving averages of 20 days. \href{https://github.com/QuantLet/SVCJrw/tree/master/SVCJrw_graph_parameters}{\protect\includegraphics[height=.9em]{qletlogo_tr.png} \color{magenta}  SVCJrw\_graph\_parameters}}
\label{svcj_param_estimates_150}
\end{figure}

Time series estimates for each parameter of the SVCJ model are presented in Figure  \ref{svcj_param_estimates_150} for the window size of 150 days. The fluctuating lines are parameter estimates, while the solid lines in their center are moving averages of 20 days. We discuss stylized patterns of representative parameters as follows.

The trend of log-return, $\mu$, moves parallelly to the CRIX (cf. Figure \ref{CRIX_plot}) as expected. Especially the growth in 2017 and the drop in 2018 are well reflected. But the drift of the volatility, $\alpha$, oscillates at a low level ($\alpha <$ 0.2) until the end of 2017, then suddenly jumps to a high level ($\alpha$ = 0.5) at the turn of 2017/2018. A similar pattern occurs at the end of 2019: less strongly, but in the same direction, $\alpha$ rises again until the end of 2020. It is worth noting that the rise in $\alpha$ correlates with downturns in the CRIX, partially because when the market is bearish, the volatility may take longer to return to its long-run trend.

The coefficient of lagged volatility, $\beta$,  also substantially correlates to the fluctuation of the CC market: before the rise in 2017, its value oscillates around $-0.4$, then it stagnates at $-0.2$ throughout 2017, though towards the end of 2017 they drop to $-0.8$. In bearish periods, the values of $\beta$ get closer to $-1$ and $\alpha$ increases, which indicates that volatility takes longer to return to its long-run level. In contrast, volatility detaches from its lagged values in times of rising markets (e.g., $\beta$ is close to zero througout 2017) and quickly returns to its long-run trend, since $\beta$ is reversely related to the mean-reversion rate.

The jump sizes ($\mu_v$ and $\mu_y$) seem to interact with the overall CC market dynamics: in bullish periods, $\mu_y$ stabilizes at low levels and the volatility of jumps in mean $\sigma_y$ stabilizes as well. But it seems that $\lambda$ doesn't reveal any remarkable pattern. \\

\subsection{Effects of Window Size}{\label{windowsize}}

\begin{figure}[h!] 
\centering
\includegraphics[width = 1.1\textwidth]{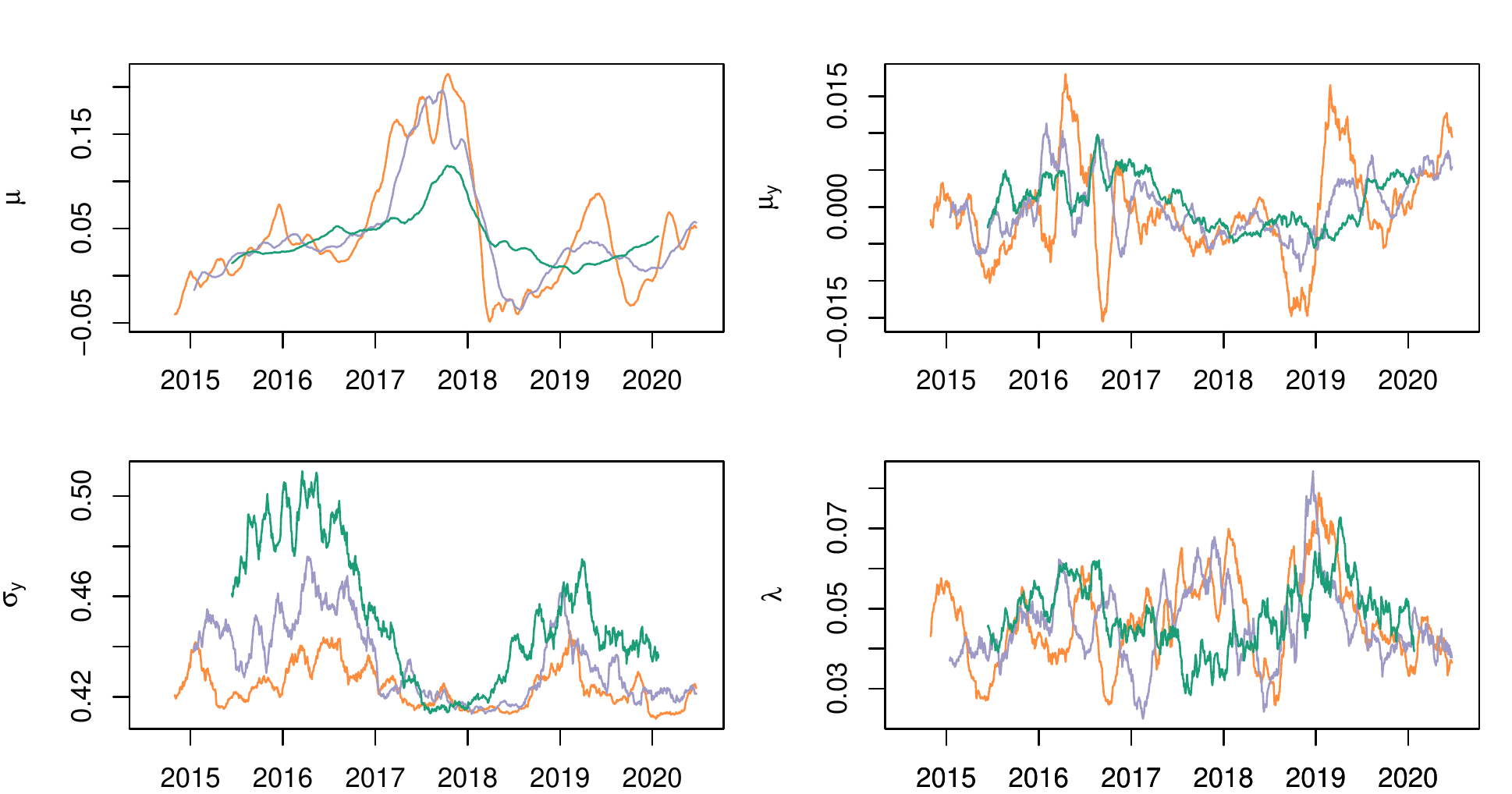}
\includegraphics[width = 1.1\textwidth]{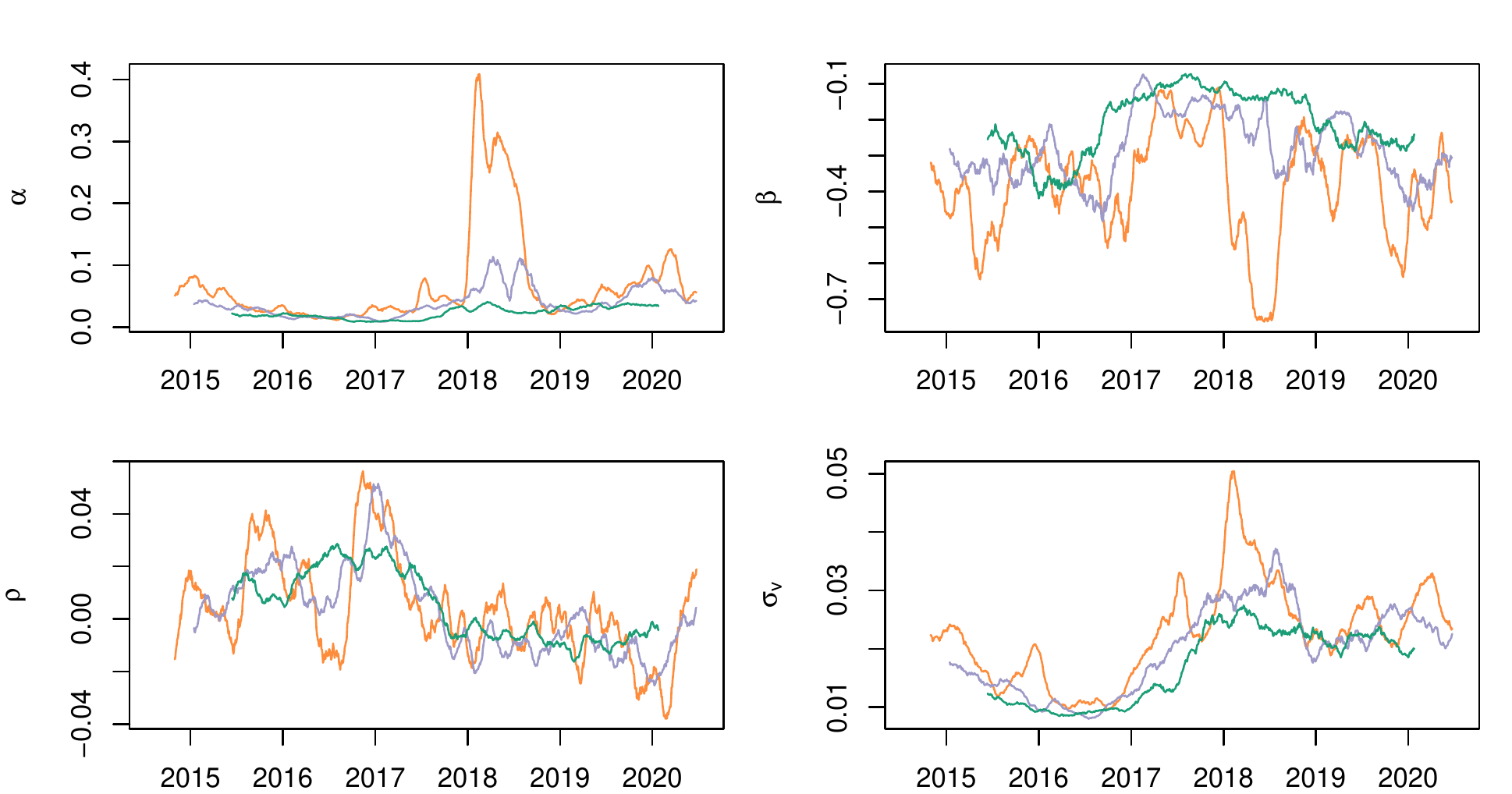}
\includegraphics[width = 1.1\textwidth]{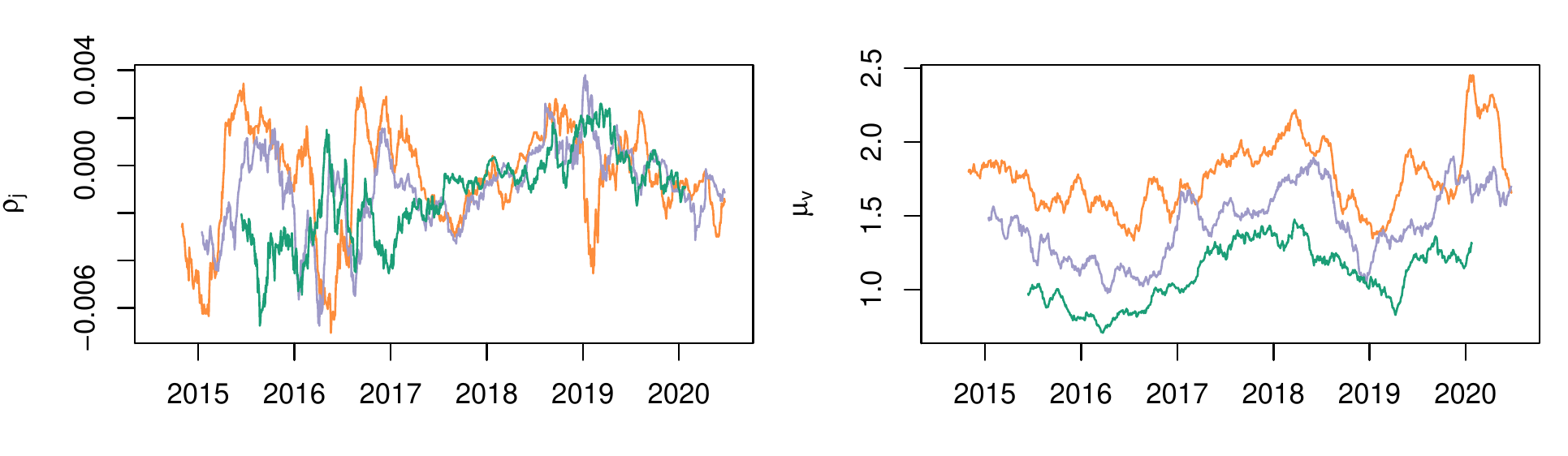}
\caption{20-day moving average of parameter estimates for several window sizes (\color{myorange}{150}, \color{my300}{300} \color{black}\& \color{my600}{600} \color{black}days). \href{https://github.com/QuantLet/SVCJrw/tree/master/SVCJrw_graph_parameters}{\protect\includegraphics[height=.9em]{qletlogo_tr.png} \color{magenta}  SVCJrw\_graph\_parameters}}
\label{svcj_param_estimates}
\end{figure}

Figure \ref{svcj_param_estimates} collects 20-day moving average estimates for several rolling windows of size 150, 300, and 600 days. It turns out that the three time series are not identical and are substantially affected by the size of the rolling windows. In general, curves are smoother for bigger window, especially protruding for the parameters $\mu, \alpha, \rho$ and $\sigma_y$. However we can still observe some consistent patterns among the three time series:

\paragraph{Volatility}
The time series of the parameters for volatility show matching dynamics. The dynamics of $\alpha$ are very robust and its three time series overlap almost over the entire period of analysis. Only around the turn of 2017/18 does $\alpha$ skyrocket, to varying degrees for each window size. This suggests that volatility is high when the market is falling.

Similarly, the time series of $\beta$ converge at a level close to Zero throughout 2017. This confirms that volatility gets detached from its lagged values when the market is bullish. By contrast, when there is no clear market direction or when the market is falling, $\beta$ deviates from Zero, i.e. volatility needs some time to return to its long-run trend and persists in its former state.

Another finding relates to the volatility of volatility $\sigma_v$: it seems that a regime change took place around the turn of the year 2017/2018. Until 2017, $\sigma_v$ fluctuated at low levels and increased strongly simultaneously with the growth of the CC sector. After 2018, volatility remained at this elevated level. Based on the CRIX time series alone, one cannot explain this level shift, but it stands to reason that the opportunities in the CC market have attracted many investors since 2018, which may have increased applications of CCs as well as speculation, thereby increasing volatility.

\paragraph{Jumps} In the previous section \ref{svcj_estimates}, we observed that the size of jumps in mean $\mu_y$ declined to zero whenever the market is rising. Interestingly, this finding can be confirmed: all three time series converge simultaneously towards zero between 2017 and 2018. And not only the size of the jumps, but also their volatility decreases rapidly when the market is rising: one can observe how the estimates of the volatility of the jumps $\sigma_y$ decrease in 2017. Weakened, but in the same direction, a convergence of the time series can be observed for the second half of 2019.

\clearpage
\subsection{Cluster Analysis}{\label{cluster_analysis}}

\begin{figure}
\centering
\hspace{0.45cm}\includegraphics[width=0.9\textwidth]{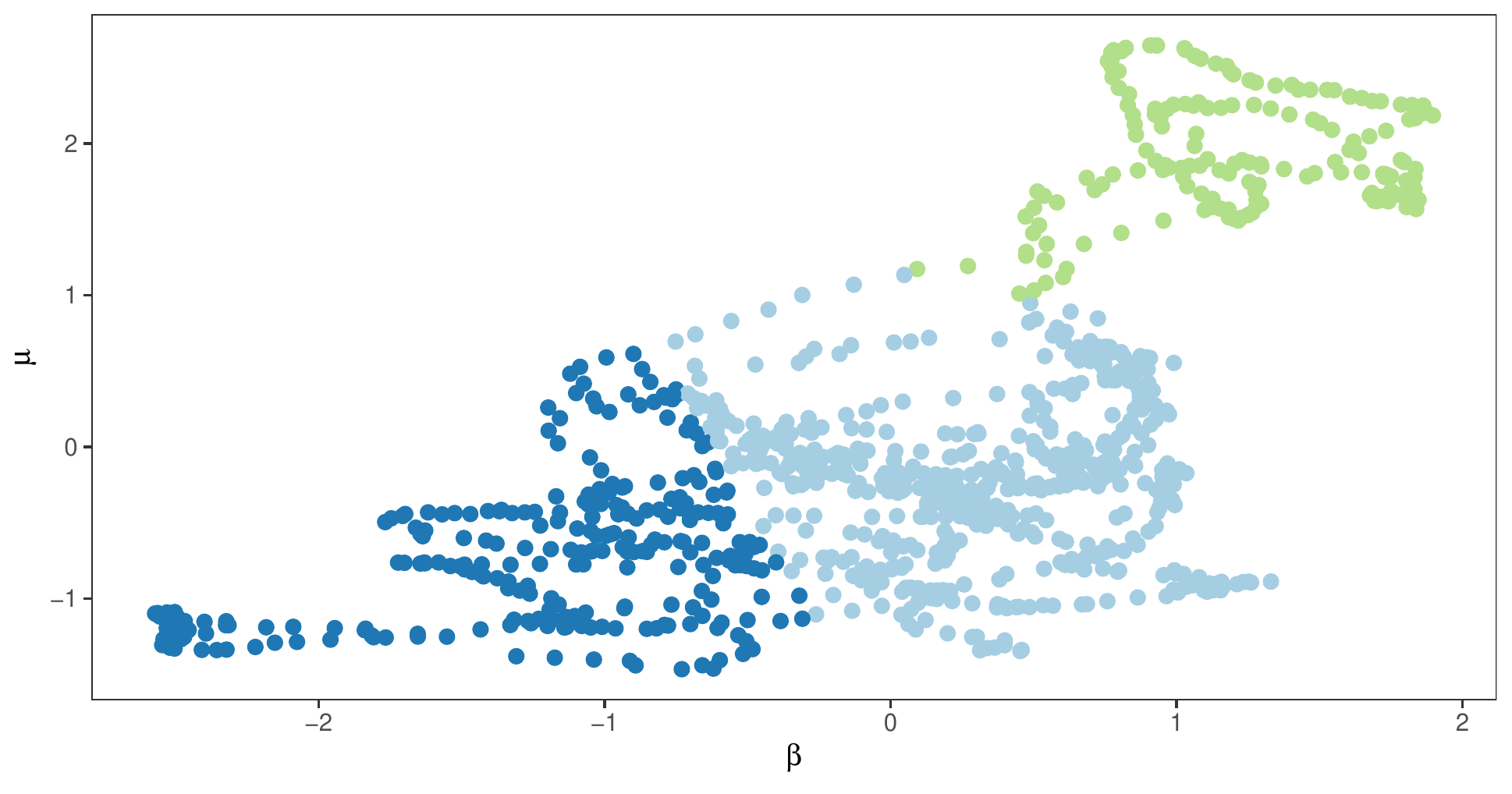}
\includegraphics[width=0.93\textwidth]{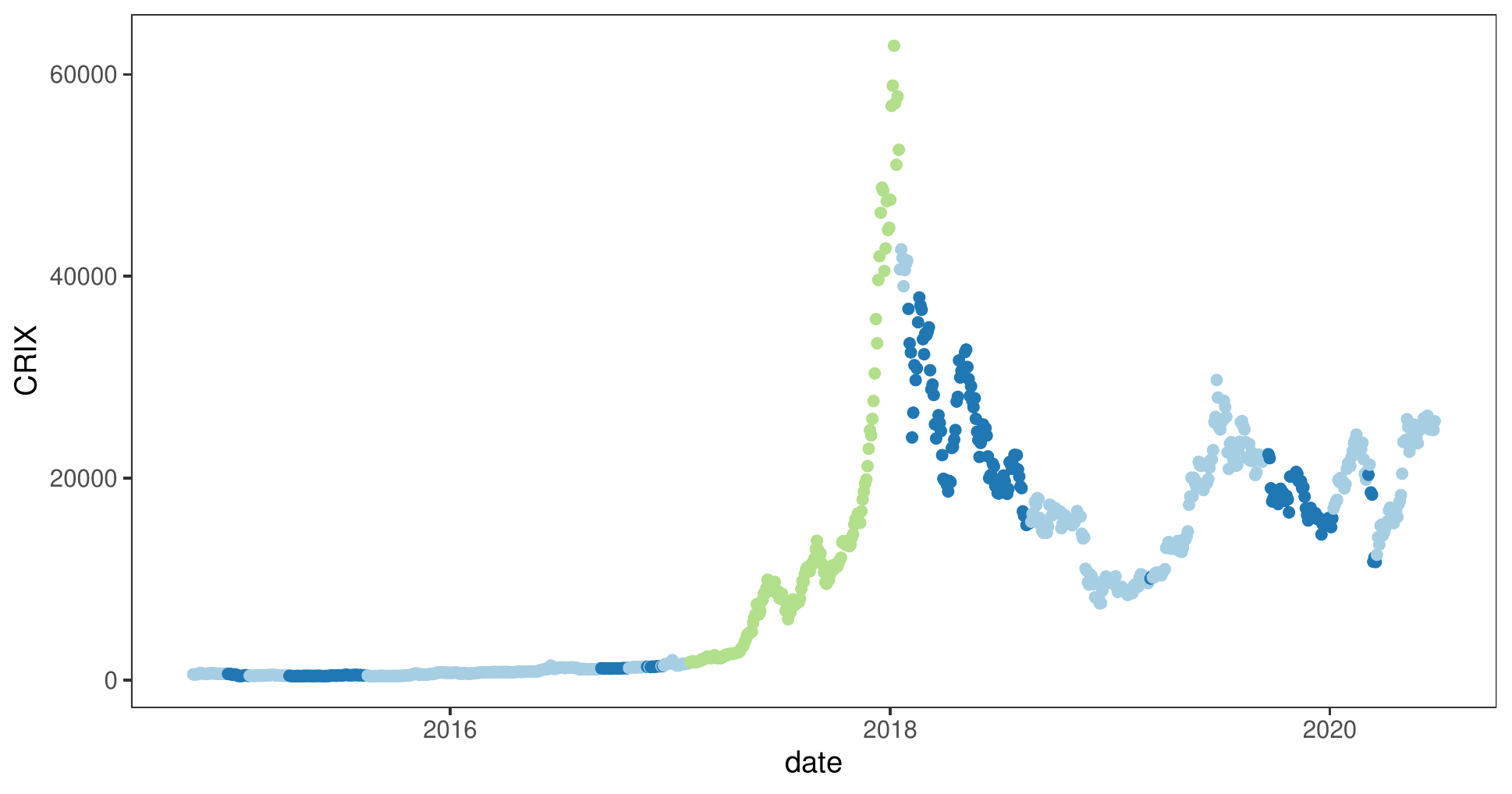}
\caption{Top: $k$-means clusters of parameter estimates $\mu$ and $\beta$, $k = 3$. \\
Bottom: the CRIX coloured by the respective clusters. \\ \href{https://github.com/QuantLet/SVCJrw/tree/master/SVCJrw_clustered_parameters}{\protect\includegraphics[height=.9em]{qletlogo_tr.png} \color{magenta}  SVCJrw\_clustered\_parameters}}
\label{crix_clustered_mu_beta}
\end{figure}

To further explore possible patterns on correlations among the parameters, we perform cluster analysis in this section.
 The clustering of parameters is not intended to show causal relationships, but merely to illustrate certain patterns. Especially when the market points in a specific direction, some parameters stabilize, that's why as first example shall serve the correlation between the trend $\mu$ and the volatility parameter $\beta$. Figure \ref{crix_clustered_mu_beta} presents $k$-means clusters for this parameter pair. The elbow method yields the optimal number of $k=3$ clusters. Below the clustered pair of parameters is the CRIX colored in the same colors as the clusters, which reveals the time dimension of the data and its link to the overall market dynamics. Note that for clustering, variables were scaled. \\
The clustering reveals interesting connections: there is a strong relationship between the two parameters $\beta$ and $\mu$. This is impressive since the underlying CRIX data is highly non-stationary. An increase in trend is accompanied by an increase in $\beta$, i.e. $\beta$ converges to zero and the current volatility breaks away from its previous values. By contrast, when the trend is decaying, $\beta$ declines as well and volatility becomes more persistent. Similar patterns have been obtained by $k$-expectiles clustering (cf. \cite{wang2021}).\\

\begin{figure}[h!]
\centering
\hspace{0.44cm}\includegraphics[width=.94\textwidth]{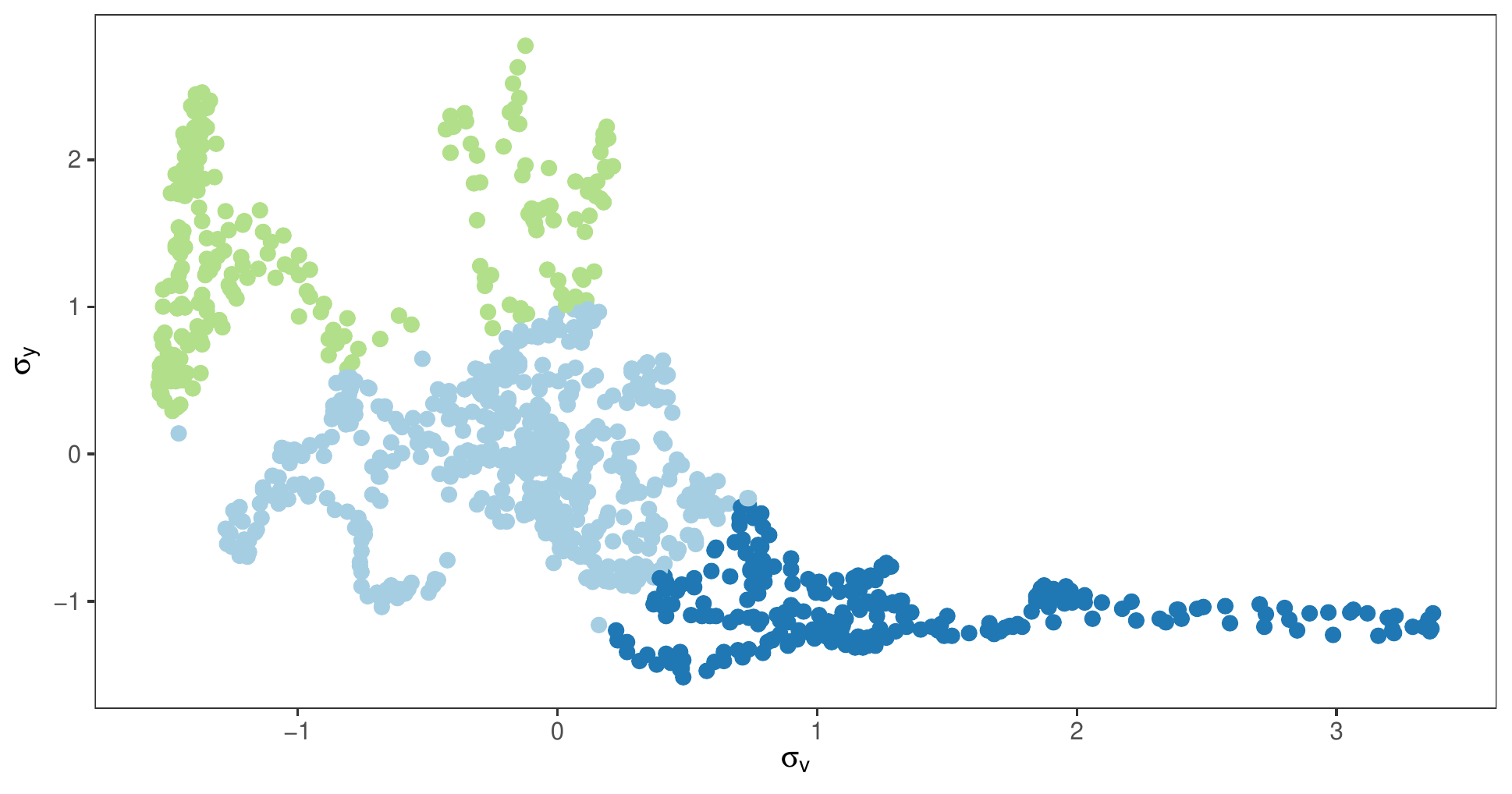}
\includegraphics[width=.98\textwidth]{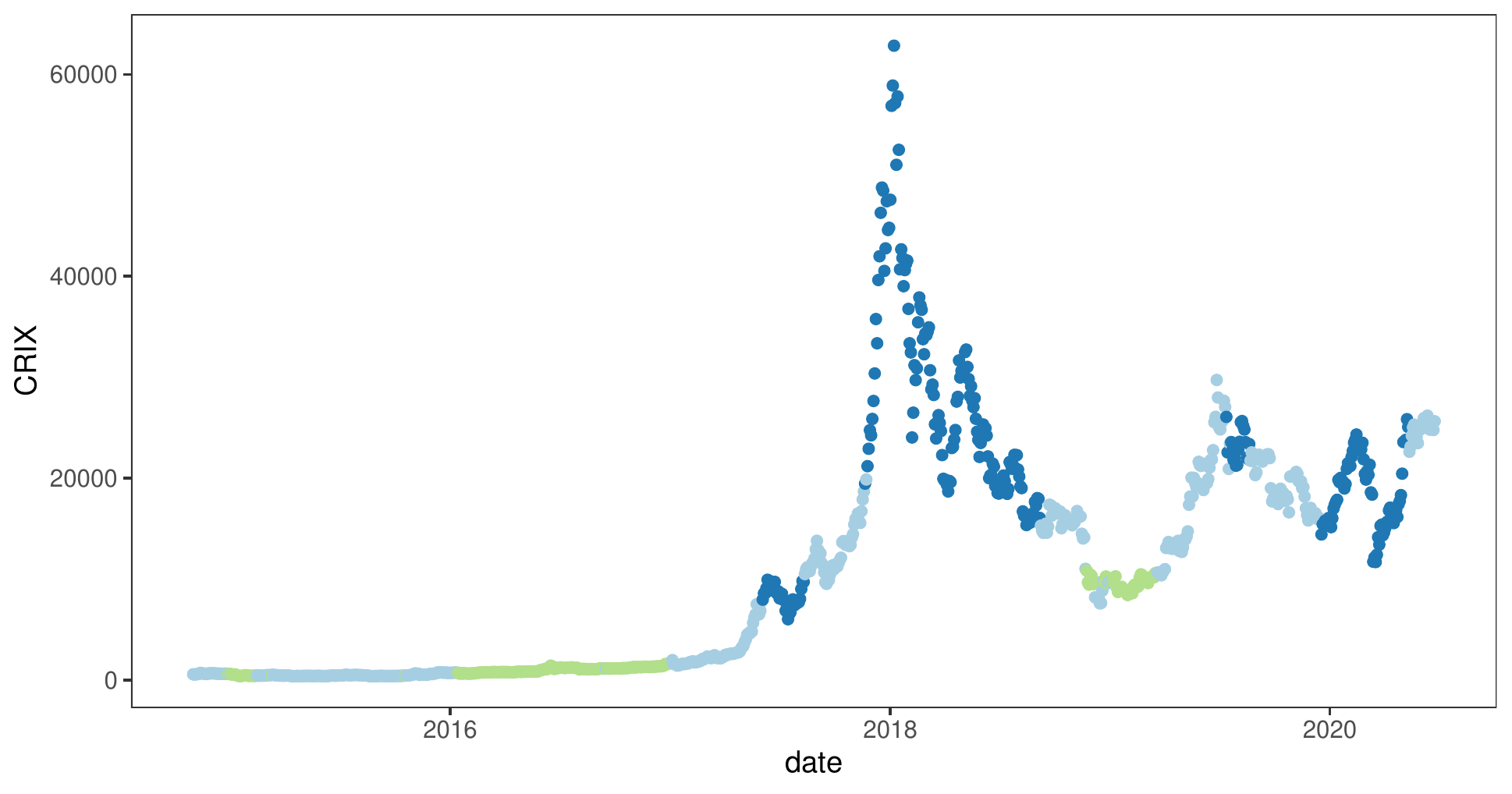}
\caption{Top: k-means clusters of parameter estimates $\sigma_y$ and $\sigma_v$, k = 3. \\
Bottom: the CRIX coloured by the respective clusters.  \href{https://github.com/QuantLet/SVCJrw/tree/master/SVCJrw_clustered_parameters}{\protect\includegraphics[height=.9em]{qletlogo_tr.png} \color{magenta}  SVCJrw\_clustered\_parameters}}
\label{cluster_sigmas2}
\end{figure}

Another example of the interactions among parameters is presented in Figure \ref{cluster_sigmas2}. The volatility of jumps in returns $\sigma_y$ and the volatility of volatility parameter $\sigma_v$ are reversely related: the volatility of jumps $\sigma_y$ is low, when the volatility of volatility parameter $\sigma_v$  increases, and vice versa. The lower graph in Figure \ref{cluster_sigmas2} illustrates the time dimension as well as the impact of the market dynamics on this correlation:
Interestingely, high volatllity of jumps in returns $\sigma_y$ occurs especially when the market is not pointing in any specific direction. In such periods, volatility of volatiliy $\sigma_v$  stays at low levels. When the market is overheated, volatility of volatility $\sigma_v$ is high. Interestingely, $\sigma_v$ started to increase before the CC sector reached its peak in 2017/18. Furthermore, we observe high volatility of volatility $\sigma_v$  and low volatility of jumps in returns $\sigma_y$ for periods of market downturns.

\section{Conclusion}
This paper examined the CC sector as a whole and shed light on its dynamics.
By combining an SVCJ model with a rolling window approach, we obtain time series for each parameter of the model. Thereby we identify several recurring patterns: first, volatility remains at a low level during bullish CC market movements and rises in times of bearish markets. In addition, when volatility is already on a high level, it needs longer to return to its long-run trend. Second, in times of bullish markets, the size of  jumps in mean return decreases, and its volatility stabilizes as well at low levels. Third, a level shift of the volatility of volatility parameter occurred simultaneously to the rise of the CC market at the turn of the year 2017/18. Finally, the jumps in mean and in volatility seem to be independent. The findings are robust to changes in the window size and confirmed by clustering of the parameters.


\bibliography{literature}

\end{document}